\documentclass[9pt,twocolumn,twoside]{osajnlKS}
\pdfoutput=1 
\usepackage{subcaption,graphicx,trimclip}
\usepackage[nameinlink]{cleveref}

\journal{ol} 
\usepackage[version=3]{mhchem}
\setboolean{shortarticle}{true}
\usepackage{cleveref}
\ifthenelse{\boolean{shortarticle}}{\colorlet{color2}{color2b}}{\colorlet{color2}{color2}}
\usepackage{hyperref}
\hypersetup{
    bookmarks=true,         
    unicode=false,          
    pdftoolbar=true,        
    pdfmenubar=true,        
    pdffitwindow=false,     
    pdftitle={Phased-Locked Two-Color Single Soliton Microcombs in Dispersion-Engineered Si3N4 Resonators},    
    pdfauthor={Dr. Gregory Moille},     
    pdfsubject={},   
    pdfcreator={Dr. Gregory Moille},   
    pdfproducer={},  
    pdfkeywords={Soliton,} {Microcombs,} {Resonators}, 
    pdfnewwindow=true,      
    colorlinks=false,       
    linkcolor=red,          
    citecolor=green,        
    filecolor=magenta,      
    urlcolor=cyan           
}

\title{Phased-Locked Two-Color Single Soliton Microcombs in Dispersion-Engineered Si\textsubscript{3}N\textsubscript{4} Resonators}
\author[1,2,*]{Gregory~Moille}
\author[1,2]{Qing~Li}
\author[1,3]{Sangsik~Kim}
\author[1]{Daron Westly}
\author[1,$\dagger$]{Kartik~Srinivasan}
\affil[1]{Center for Nanoscale Science and Technology, National Institute of Standards and Technology, Gaithersburg, Maryland 20899, USA}
\affil[2]{Maryland NanoCenter, University of Maryland, College Park, Maryland 20742, USA}
\affil[3]{Department of Electrical and Computer Engineering, Texas Tech University, Lubbock, TX 79409 USA}
\affil[*]{gregory.moille@nist.gov}
\affil[$\dagger$]{kartik.srinivasan@nist.gov}

\dates{\today}
\ociscodes{(140.3498) Microcavity devices; (190.4390) Nonlinear optics, integrated optics; (190.3270) Kerr effect}
\doi{\url{http://dx.doi.org/10.1364/ao.XX.XXXXXX}}

\begin{abstract}
We propose and theoretically investigate a dispersion-engineered Si\textsubscript{3}N\textsubscript{4} microring resonator, based on a cross-section containing a partially-etched trench, that supports phase-locked, two-color soliton microcomb states.  These soliton states consist of a single circulating intracavity pulse with a modulated envelope that sits on a continuous wave background.  Such temporal waveforms produce a frequency comb whose spectrum is spread over two widely-spaced spectral windows, each exhibiting a squared hyperbolic secant envelope, with the two windows phase-locked to each other via Cherenkov radiation.  The first spectral window is centered around the 1550 nm pump, while the second spectral window is tailored based on straightforward geometric control, and can be centered as short as 750 nm and as long as 3000 nm.  We numerically analyze the robustness of the design to parameter variation, and consider its implications to self-referencing and visible wavelength comb generation.
\end{abstract}

\graphicspath{{Figure/}}

\begin{document}

\maketitle
\thispagestyle{fancy}

\ifthenelse{\boolean{shortarticle}}{\ifthenelse{\boolean{singlecolumn}}{\abscontentformatted}{\abscontent}}{}

\begin{figure}[t]
    \begin{center}
        \includegraphics{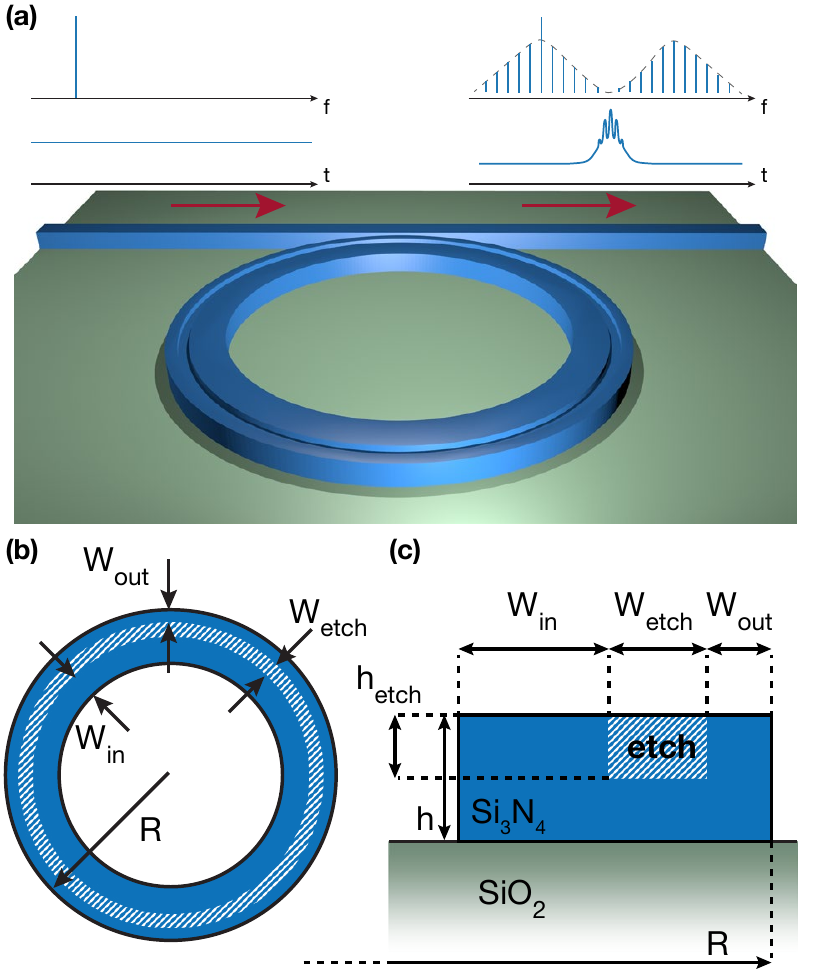}
        \caption{\label{fig:DesignParameters} Slot ring resonator design parameters on \ce{SiO2} substrate with air top cladding. (a) Schematic of the trench-ring geometry, producing from a continuous wave (CW) input a two-color single soliton state.  The spectrum and temporal waveform associated with the input and output are indicated above the schematic diagram. (b) Top view, (c) cross section of the resonator. Blue color corresponds to Si\textsubscript{3}N\textsubscript{4}, the hatched part to the etched slot.}
    \end{center}
    \vspace{-4ex}
\end{figure}

\vspace{-2ex}
\noindent Optical frequency synthesis~\cite{holzwarth_optical_2000}, low-noise microwave generation~\cite{fortier_generation_2011}, and optical atomic clocks~\cite{ludlow_optical_2015} often rely on a fully-stabilized optical frequency comb as a core component~\cite{diddams_evolving_2010}. Stabilization of the comb's carrier-envelope offset frequency is typically achieved via nonlinear interferometry, such as $f$-2$f$ self-referencing~\cite{udem_optical_2002,diddams2000direct,telle1999carrier}.  Optical frequency combs based on Kerr nonlinear microresonators~\cite{kippenberg_microresonator-based_2011} have recently emerged as a compelling approach to realizing compact, low-power operation, with octave-spanning soliton microcombs~\cite{li_stably_2017,pfeiffer_octave-spanning_2017} having recently been self-referenced~\cite{briles_kerr-microresonator_2017,spencer2017integrated}.  Despite these recent advances, direct self-referencing of Kerr microcombs is quite challenging, with experimental demonstrations to date having required helper/transfer lasers, phase locked to a nearby comb tooth, to achieve detectable offset frequency beat notes~\cite{briles_kerr-microresonator_2017,spencer2017integrated,brasch2017self}. The use of such lasers comes at the expense of additional system complexity, component cost, and operating power. A fundamental challenge in self-referencing a Kerr microcomb is that solitons have thus far been the only states to experimentally exhibit both the requisite broad spectral bandwidth and phase coherence across the comb, but the soliton's characteristic squared hyperbolic secant spectral envelope limits the power in the comb lines relevant for self-referencing. Soliton-induced dispersive wave emission~\cite{brasch_photonic_2016,briles_kerr-microresonator_2017} has been identified as a means to dramatically enhance the power available in the relevant comb teeth, but the ideal scenario of achieving precise control of two dispersive waves (preferably at $f$ and 2$f$) is a significant challenge~\cite{Li15}, in particular due to their narrow spectral width and the sensitivity of the phase-matching conditions to geometric parameters~\cite{cherenkov_dissipative_2017}.  In addition, developing soliton microcombs to cover other spectral windows, for example, to connect to visible wavelength atomic systems for clocks or to longer wavelengths for spectroscopy, often requires a wavelength change in the pump laser source. %
To overcome such challenges, the generation of multi-color single soliton states as proposed in Ref.~\cite{luo_multicolor_2016} is of great interest. These states (Fig.~\ref{fig:DesignParameters}(a)) produce a spectrum that is broken up into multiple windows, each of which exhibits a $sech^2$ envelope.  Temporally, they still consist of a single pulse circulating through the cavity, sitting on a continuous wave background, but their envelope exhibits modulations consistent with the multiple central frequencies . An open question is the degree to which realizable devices can exhibit the necessary dispersion profile to support such states, and the extent to which the dispersion can be optimized for specific applications.

Here, we present the theoretical analysis of a fabrication-friendly design for the generation of two-color soliton states, based on a straightforward modification to Si\textsubscript{3}N\textsubscript{4} resonators that have successfully exhibited octave-spanning (single color) soliton states~\cite{li_stably_2017}. Each soliton color is supported by a local region of anomalous group velocity dispersion (GVD), and phase-matching of the two regions phase locks the two soliton colors to each other. Simple geometrical changes keep one soliton color centered at 200~THz (1550~nm) while the position of the second color is widely varied from 100~THz (3000~nm) to 400~THz (775~nm), with comb teeth at wavelengths as short as 680~nm. As we describe below, this is an appealing approach for the realization of self-referenced microresonator frequency combs that can address multiple relevant spectral regions, all while being pumped by a telecommunications-band laser.

\begin{figure}[t]
    \begin{center}
        \includegraphics{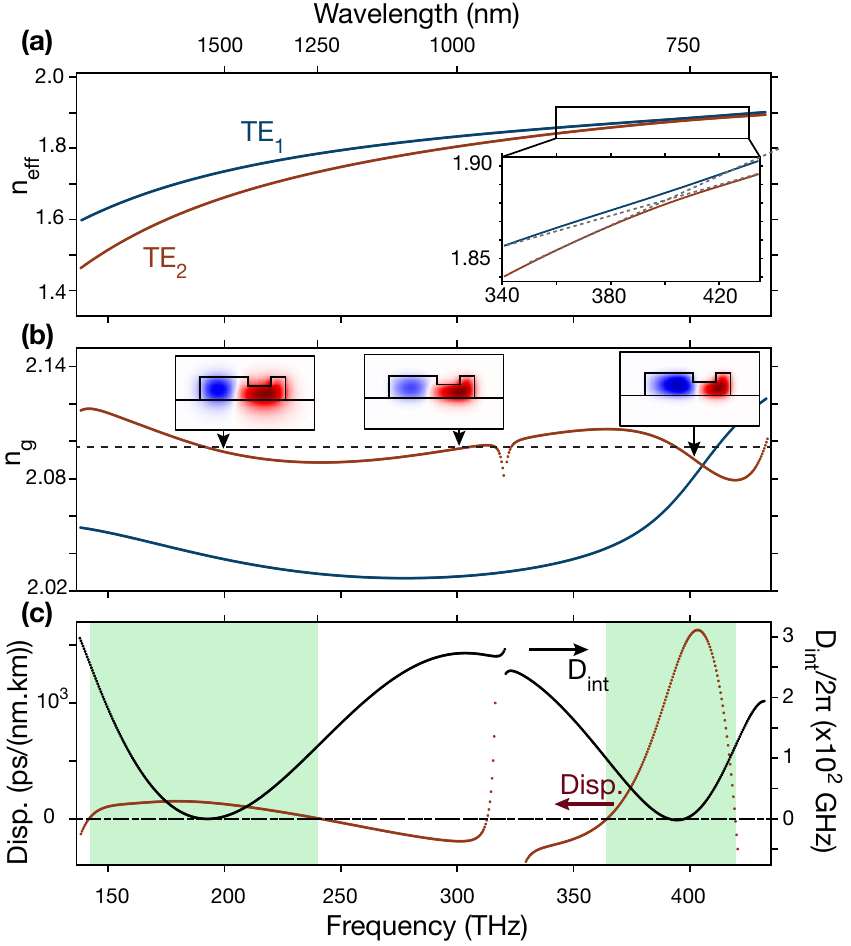}
        \caption{\label{fig:DispersionNominal} (a) Effective index versus frequency of the fundamental TE mode (TE\textsubscript{1}, blue) and second TE mode (TE\textsubscript{2}, red). The inset corresponds to the frequency position of the avoided mode-crossing. (b) Group index of TE\textsubscript{1} and TE\textsubscript{2} (blue and red respectively). The radial component of the mode profile is displayed at the selected frequency for the TE\textsubscript{2} mode. (c) Dispersion (red, left scale) and integrated dispersion (black, right scale) for the TE\textsubscript{2} mode. The green areas highlight the two anomalous dispersion regions for TE\textsubscript{2}.}
    \end{center}
    \vspace{-5ex}
\end{figure}
To support a two-color soliton state as schematically depicted in Fig.~\ref{fig:DesignParameters}(a), a Kerr microresonator must support a GVD profile that contains two anomalous GVD regions (and hence four zero-dispersion crossing points), each of which sustains a color of the soliton, and the two anomalous GVD windows must be phase-matched to allow energy exchange between the two soliton colors. As the basic rectangular (or trapezoidal) cross-section used in broad soliton microcombs to date only results in one anomalous GVD window, a more complex geometry must be introduced. A number of groups have considered such complex dispersion engineering, with most approaches relying on a layered cross section consisting of materials such as Si\textsubscript{3}N\textsubscript{4}, SiO$_2$, and SiC~\cite{boggio_dispersion-optimized_2016,boggio_dispersion_2014,jafari_fabrication-friendly_2016,liang_ultra-broadband_2016}. However, such an approach introduces additional challenges in fabrication, in requiring the deposition, and in some cases, etching, of additional materials with fine control of dimension. In particular, we have found that the thickness control required to generate an appropriate dispersion profile for two-color solitons in the Si\textsubscript{3}N\textsubscript{4}/SiO$_2$ platform is quite strict.  Our proposed geometry, depicted in Fig.~\ref{fig:DesignParameters}, instead relies on the straightforward etching of a trench within an air top clad, SiO$_2$ bottom clad, Si\textsubscript{3}N\textsubscript{4} microring resonator (radius $R=50$ $\mu$m and thickness of $h=700$ nm). Nominal dimensions for the trench are width $W_{etch} = 775$ nm, depth $h_{etch} = 275$ nm, and position from the outer (inner) part of the ring of $W_{out}=475$ nm ($W_{in}=1.8$ $\mu$m).

The trench is the key enabling feature for the requisite dispersion engineering, in enabling control of the spectral position and strength of an avoided mode crossing between the two first sets of transverse electric (TE) modes, namely TE\textsubscript{1} and TE\textsubscript{2}. The effective index, group index, and dispersion of the these two modes are computed (Fig.~\ref{fig:DispersionNominal}(a)-(c)) using a finite-element method eigensolver that includes material dispersion models for \ce{Si3N4} and SiO$_2$. Both modes exhibit a first anomalous dispersion region around 200~THz, due to the compensation of the material dispersion by the ring dispersion (combination of waveguiding and bending effects). An avoided mode crossing around 400~THz (Fig.~\ref{fig:DispersionNominal}(a)) leads to a second anomalous dispersion region for the TE$_2$ mode around this frequency (Fig.~\ref{fig:DispersionNominal}(b)-(c)). This is similar to the recently reported anomalous dispersion generated by the avoided mode crossing between symmetric and anti-symmetric modes in concentric coupled resonators~\cite{kim_dispersion_2017}. In addition, the two regions match both group and phase velocity (Fig.~\ref{fig:DispersionNominal}(b)-(c)). Indeed,the integrated dispersion, defined by $D_{int} = \omega_\mu - (\omega_0 + D_1 \mu$), where $\mu$ is an integer representing the relative mode number from the pumped one, $\omega_0$ is the he angular frequency of the pumped mode, $\omega_{\mu}$ the angular frequency of the $\mu$\textsuperscript{th} mode, and $D_1$ the free spectral range of the resonator, is zero for both the pumped mode and the central frequency of the anomalous dispersion region created by the avoided mode crossing. This allows energy transfer through Cherenkov radiation~\cite{akhmediev_cherenkov_1995} between the two regions. Hence, by pumping one anomalous dispersion region and creating a first soliton color, power is transferred to the other anomalous dispersion region, ultimately leading to the formation of a second soliton color~\cite{luo_multicolor_2016}.

\begin{figure}[t]
    \begin{center}
        \includegraphics{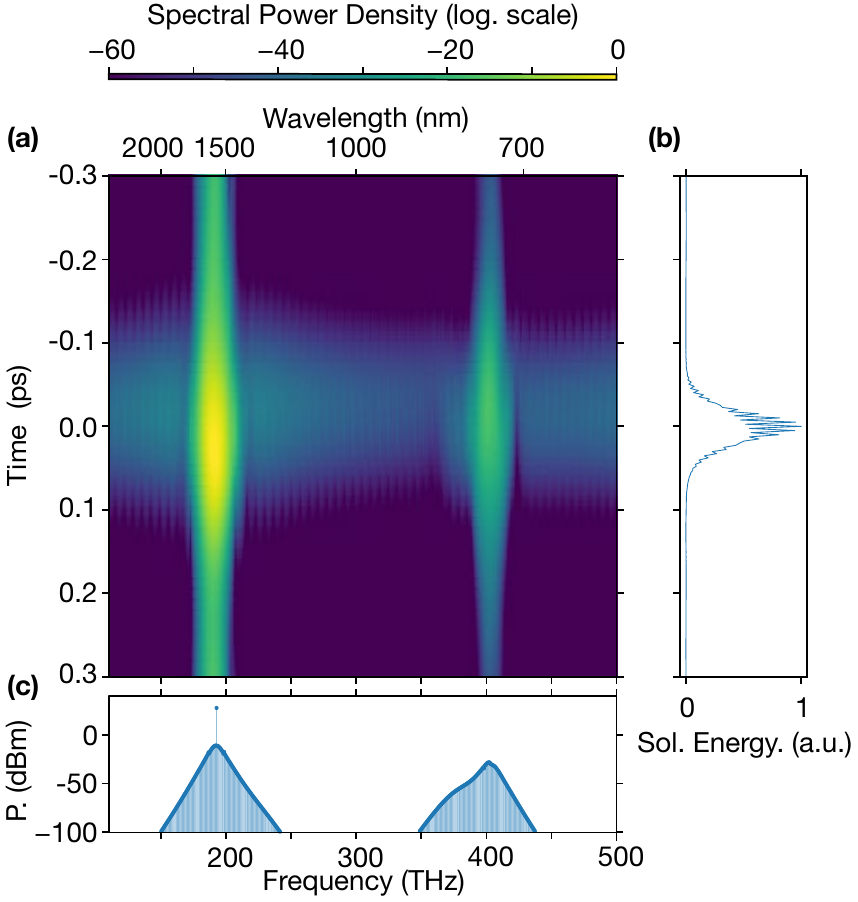}
        \caption{\label{fig:200THz400THzSpectrogram} (a) Spectrogram of the two-color soliton (log-scale color map). (b) Temporal profile of the multicolor soliton. (c) Output spectrum of the two-color soliton; 0 dB is referenced to 1 mW (i.e., dBm).}
    \end{center}
    \vspace{-5ex}
\end{figure}

To investigate frequency comb generation in the soliton regime of this two matched anomalous GVD design, the Lugiato-Lefever equation~\cite{coen_universal_2013}, taking into account the dispersion of the TE$_2$ mode previously computed, is solved in the temporal domain and results in the bound two-color soliton frequency comb represented in Fig.~\ref{fig:200THz400THzSpectrogram}(a)-(c). Here we use a coupling and intrinsic quality factor $Q_c = Q_i = 6\times10^6$ and a pump located in the C-band with a power of $|E_{in}|^2 =600$~mW. The obtained spectrum exhibits two characteristic $sech^2$ envelopes at the positions of the two anomalous dispersion regions, namely 200~THz and 400~THz (Fig.~\ref{fig:200THz400THzSpectrogram}(c)). An important aspect to point out is that the two colors of the comb are part of a single soliton pulse, and are thus temporally bound and phase-locked, as represented in the spectrogram in Fig.~\ref{fig:200THz400THzSpectrogram}(a). The result is a temporal single soliton pulse with a modulated envelope (Fig.~\ref{fig:200THz400THzSpectrogram}(b)) due to the beating between the two spectral comb regions $|S(t)|^2 \propto  P_1 \mathrm{sech}^2\left( t/\Delta\tau_1\right) +P_2 \mathrm{sech}^2\left( t/\Delta\tau_2\right) + 2\sqrt{P_1P_2}  \mathrm{sech}\left( t/\Delta\tau_1\right) \mathrm{sech}\left( t/\Delta\tau_2\right) \mathrm{cos}\left(\left(\omega_1 - \omega_2\right)t\right)$, where $\omega_{\{1,2\}}$ are the central frequencies of the two frequency combs, and $\Delta\tau_{\{1,2\}}$ are the temporal widths of the uncoupled solitons generated by each anomalous dispersion region with peak power $P_{\{1,2\}}$.



\begin{figure}[t]
    \begin{center}
        \includegraphics{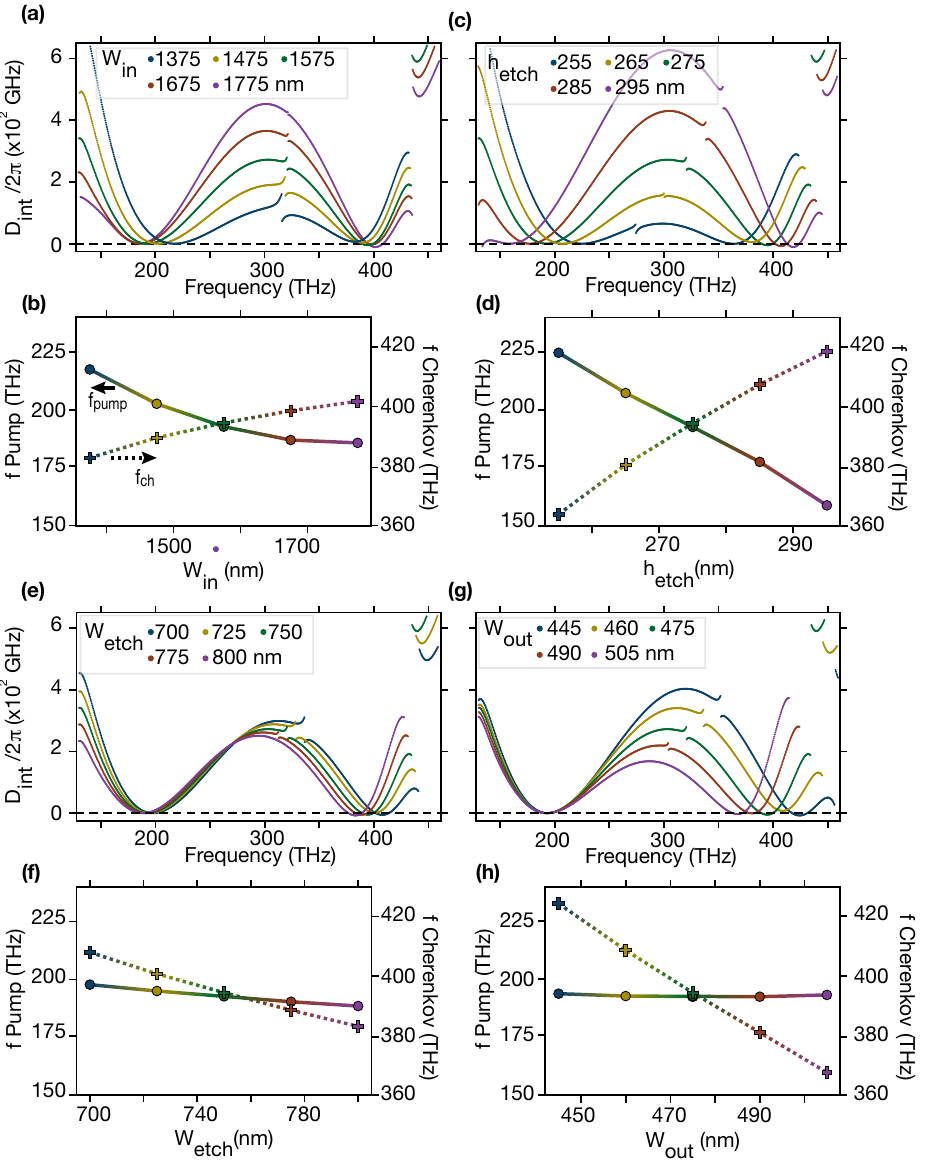}
        \caption{\label{fig:Robustness}
        Robustness of the design for generating two-color solitons with regards to: (a)-(b) the inner width $W_{in}$, (c)-(d) the etch depth $h_{etch}$, (e)-(f) the etch width $W_{etch}$ and (g)-(h) the outer width $W_{out}$. Figures (a), (c), (e), and (g) are the integrated dispersion for several values of the investigated parameter; (b), (d), (f), and (h) are the shift of phase-matched pump (solid line, left y-axis) and Cherenkov radiation frequencies (dashed line, right y-axis).}%
    \end{center}
    \vspace{-5ex}
\end{figure}

\begin{figure*}[t]
    \begin{center}
        \includegraphics{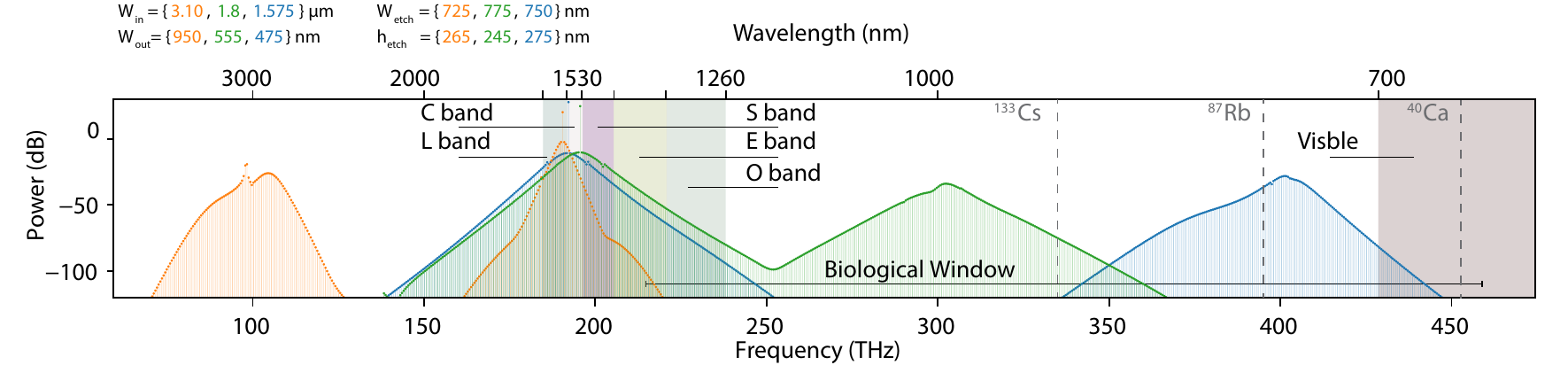}
        \caption{\label{fig:CompareSpectra}Spectra of the bound two-color soliton frequency comb for different design parameters tuning the center of the second soliton color at 100~THz (orange), 300~THz (green) and 400~THz (blue) while still pumping around 200~THz; 0 dB is referenced to 1 mW (i.e., dBm).}
    \end{center}
    \vspace{-4ex}
\end{figure*}


While it is important to point out the fabrication simplicity of our design, we need to ensure that the two-color soliton behavior is reasonably robust with respect to expected fabrication tolerances. Hence, simulations are performed where all the parameters from Fig.~\ref{fig:DesignParameters} are tuned independently, from which we extract the integrated dispersion and the position of both the pumped and the Cherenkov frequencies where the phase matching condition is respected, which in first order approximation corresponds to where $D_{int} = 0$. The least sensitive parameter is the inner width (Fig.~\ref{fig:Robustness}(a)-(b)), which is consistent with the fact that it corresponds to the biggest component of the trenched microring, while the most sensitive parameter is the etch depth (Fig.~\ref{fig:Robustness}(c)-(d)) as a variation of a few percent of the dimension could lead to a change in the central positions of both anomalous GVD windows. However, even with a variation of $\pm20$~nm from the nominal etch depth (i.e., consistent with the expected fabrication uncertainty), the ability to achieve two phase-locked anomalous GVD regions remains. Interestingly, the width of the etched trench (Fig.~\ref{fig:Robustness}(e)-(f)), has only a limited influence on the spectral positions of the two anomalous GVD regions. As pointed out in  ref.~\cite{kim_dispersion_2017}, this dimension tunes the coupling between TE$_1$ and TE$_2$. Thus, for a narrower trench (i.e. stronger coupling), the integrated dispersion in between the two anomalous GVD regions is higher than for a wider trench (i.e. weaker coupling), as the avoiding mode crossing is steeper. Finally, it is worth pointing out that the outer width (Fig.~\ref{fig:Robustness}(g)-(h)) allows one to tune the frequency of the Cherenkov radiation without modifying the pump frequency.

Given the possibility shown above to modify the frequency position of the Cherenkov radiation without having to strongly modify the pump wavelength, we show here that this trench design allows tuning of the frequency position of the mode crossing, and thus engineering the position of the second anomalous dispersion region to suit the application. Figure~\ref{fig:CompareSpectra} shows the two-color soliton frequency comb spectra for different geometric parameters. The dispersion of the TE\textsubscript{2} mode is computed and the resulting soliton frequency combs are simulated using the temporal LLE as discussed above. The same intrinsic and coupling quality factor and the same pump power as the nominal design producing the reported 200~THz/400~THz two-color soliton comb in Fig.~\ref{fig:200THz400THzSpectrogram} are used. While the directly pumped soliton color remains centered near 200~THz in all the cases studied, the second color central frequency goes from from 100~THz to more than 400~THz by simple geometrical tuning of different parameters of the trenched microring. For the 100~THz second color comb, the pump is situated in the anomalous GVD region created by the avoided mode crossing while for the other cases the pump lies in the anomalous GVD region created by the compensation of the material dispersion by the ring. The second color of the soliton comb can be tailored to reach any value between these two extreme cases through appropriate choice of the geometrical dimensions. As noted in the figure, this two-color soliton approach enables a direct connection between the telecommunications bands and several wavelengths of interest for atomic spectroscopy and clocks.

This design has exciting potential in the ability to self-reference microcombs. As we have shown, the two-color soliton frequencies can be harmonic (e.g., 200~THz/400~THz or 200~THz/100~THz), so that in contrast to recent octave-spanning soliton microcombs~\cite{li_stably_2017,pfeiffer_octave-spanning_2017}, the pump can directly take part in self-referencing.  As the pump can be $\approx$20~dB stronger than any other comb line in a soliton, this is of significant benefit, particularly in the 200~THz/400~THz case, where the pump would be frequency doubled. Compared to dispersive wave emission, use of a second soliton color in self-referencing provides a much more slowly decaying spectral envelope, easing the strict fabrication requirements needed to achieve harmonic dispersive wave emission. Finally, we anticipate that increasing the number of soliton colors (e.g., three-color or four-color solitons) should be possible by including multiple trenches within the resonator cross-section.
\vspace{-2ex}
\section*{Funding}
\vspace{-1ex}
We acknowledge partial funding support from the DARPA ACES and DODOS programs.
\vspace{-2ex}
\section*{Acknowledgment}
\vspace{-1ex}
K.S. thanks Qiang Lin for helpful discussions in introducing him to the concept of multi-color solitons. G.M., Q.L., and S.K. acknowledge support under the Cooperative Research Agreement between the University of Maryland and NIST-CNST (award no. 70NANB10H193).
\vspace{-2ex}
{\bibliographystyle{osajnl}
\bibliography{MultiColorSoliton}
}

\end{document}